\documentclass [prb,twocolumn,a4paper,showpacs]{revtex4}
\usepackage{graphics}

\begin{document}

\title {Hartree-Fock ground state of the two-dimensional electron
gas with Rashba spin-orbit interaction}

\author {L.\ O.\ Juri}
\author {P.\ I.\ Tamborenea}
\affiliation{
\mbox{Departamento de F{\'{\i}}sica ''J.\ J.\ Giambiagi'',
Universidad de Buenos Aires}
\mbox{Ciudad Universitaria, Pabell\'on I,
C1428EHA Ciudad de Buenos Aires,} 
Argentina \\
\emph{Submitted 26 February 2007, published in Phys.\ Rev.\ B 27 June 2008}
}

\noindent

\begin{abstract}
We search for the uniform Hartree-Fock ground state of the
two-dimensional electron gas formed in semiconductor
heterostructures including the Rashba spin-orbit interaction.
We identify two competing quantum phases: a ferromagnetic one with
partial spin polarization in the perpendicular direction and a
paramagnetic one with in-plane spin. We present a phase
diagram in terms of the relative strengths of the Rashba to the
Coulomb interaction and the electron density. We compare our
theoretical description with existing experimental results obtained
in GaAs-AlGaAs heterostructures.
\end{abstract}

 \maketitle

\section{\label{sec:intro}introduction}

The two-dimensional electron gas (2DEG) is a paradigmatic system of
semiconductor physics and technology. Traditionally, the electronic
spin degree of freedom in this system has played a secondary role.
This situation has changed recently with the emergence of the
promising field of spintronics.\cite{aws-los-sam,zut-fab-das} The
most controllable and often predominant spin-orbit coupling in
semiconductor 2DEGs is the Rashba interaction.\cite{ras,win} It is
then important to determine the various many-body properties of the
2DEG in its presence. The ground state of the uniform 2DEG without
spin-orbit interaction is not known exactly.\cite{cep,ortiz,att} In
this Brief Report we concentrate on how the single-particle Rashba
term affects the ground state in the Hartree-Fock (HF) mean-field
approximation.\cite{gro-run-hei} While for the uniform electron gas
without spin-dependent potentials, the HF approach yields trivial
single-particle spin-orbitals (plane waves and pure spin states),
the presence of the Rashba spin-orbit interaction causes the HF
solution to possess an intriguing spin texture in momentum space. In
this Brief Report we formulate the HF theory and obtain solutions
for the uniform 2DEG with Rashba interaction. Our main finding is a
spatially uniform ferromagnetic phase, characterized by a net
out-of-plane partial magnetization appearing in a window of
densities.\cite{che-sim-giu}

\section{\label{sec:forma}hartree-fock theory with Rashba spin-orbit interaction}

The Hamiltonian of the 2DEG in the presence of the Rashba spin-orbit
interaction is
%
%
\begin {equation}
H = \sum_{i}H_{R,i}
    + \frac{1}{2}
      \sum_{i\neq j}
         v(\mathbf{r}_i - \mathbf{r}_j).
\label{eq:htotal}
\end {equation}
The second term gives the interparticle Coulomb interaction
$v(\mathbf{r}_i-\mathbf{r}_j)=  e^{2} / \varepsilon
|\mathbf{r}_i-\mathbf{r}_j|$, where $\varepsilon$ is the dielectric
constant of the semiconductor, and the first term is the sum of the
Rashba Hamiltonians of the individual electrons. The latter are
given by
%
%
\begin{equation}
H_R= -\frac{\hbar^{2}\nabla^{2}}{2m^{*}}
         -i\alpha \langle E_{z}\rangle
        \left( \sigma_{x} \frac{\partial}{\partial y}
              -\sigma_{y} \frac{\partial}{\partial x}
        \right).
\label{eq:hrashba}
\end{equation}
Here, $m^*$ is the conduction-band effective mass, $\alpha \langle
E_{z}\rangle$ is a structural parameter that determines the strength
of the Rashba coupling, and $\sigma_{x}$ and $\sigma_{y}$ are Pauli
matrices. The Rashba Hamiltonian can be solved analytically, and the
following wave functions and energies \mbox{are obtained:
\cite{ras}}
%
%
\begin {equation}
\psi_{\mathbf{k}s}(\mathbf{r})= \frac{1}{\sqrt{2A}}
                                e^{i\mathbf{k}\cdot\mathbf{r}}
\left(
      \begin{array}{c}
          i s e^{-i \varphi} \\
          1
      \end {array}
\right),
\label{eq:spinorash}
\end {equation}
%
%
\begin {equation}
  E(k,s) = \frac{\hbar^{2} k^{2}}{2m^{*}}
           + s \alpha \langle E_{z} \rangle k.
\label{eq:energirash}
\end {equation}

In these expressions $\mathbf{k}$ is the two-dimensional wave
vector, $\varphi$ is the angle of $\mathbf{k}$ in polar coordinates,
and $A$ is the surface area of the sample. The spin quantum number
$s = \pm 1$ denotes spin-up and spin-down eigenstates with respect
to the spin quantization axis, which lies in the x-y plane and is
perpendicular to $\mathbf{k}$ with a polar angle
$\phi_{R}(\mathbf{k})=\varphi -\pi /2$. Notice that, however, the
Rashba ground state is paramagnetic.\cite{win} Also, the Rashba
Hamiltonian $H_R$ is time-reversal invariant. This invariance
requires that $\psi_{\mathbf{k}s}(\mathbf{r})$ and
$\psi_{\mathbf{-k}s}(\mathbf{r})$ be Kramers-conjugate states with
the same energy eigenvalue $E(k,s)$ (Ref.\ \onlinecite{rashba}).

For a spatially uniform solution the HF spin-orbitals can be written
as
%
%
\begin {equation}
   \psi_{\mathbf{k}s}(\mathbf{r})=
     \frac{1}{\sqrt{2A}}e^{i\mathbf{k} \cdot \mathbf{r}}\chi(\mathbf{k},s).
\label{eq:spinorhf}
\end {equation}
The spinor $\chi(\mathbf{k},s)$ has components
$\chi_{\pm}(\mathbf{k},s)$, which are the unknown spinor amplitudes
to be determined through the HF procedure. Notice that while the
Rashba amplitudes of Eq.\ (\ref{eq:spinorash}) depend only on the
polar angle $\varphi$ of $\mathbf{k}$,  we allow
$\chi_{\pm}(\mathbf{k},s)$ to depend also on the modulus of
$\mathbf{k}$. The spin quantum number $s = \pm 1$ denotes, like in
the noninteracting Rashba problem, the up- and down-spin eigenstates
in an unknown spin-quantization axis $\hat{\mathbf{u}}(\mathbf{k})$
with polar angles $\theta(\mathbf{k})$ and $\phi(\mathbf{k})$.

The functional that one has to minimize is given by (direct Coulomb
terms drop out in the jellium-model electron gas)
%
%
\begin{eqnarray}
&& \mathcal{F} = \sum_{s,i}^{N_{s}}
      \chi^{\dagger}(\mathbf{k}_{i},s)
      [H_{R}(\mathbf{k}_{i}) - E(\mathbf{k}_{i},s)]
      \chi(\mathbf{k}_{i},s) \nonumber   \\
&&      -\frac{1}{2A}
      \sum_{ss',ij}^{N_{s}N_{s'}}
             v(\mathbf{k}_{i}-\mathbf{k}_{j})
             [\chi^{\dagger}(\mathbf{k}_{j},s')
             \chi(\mathbf{k}_{i},s)]  \nonumber   \\
&&           \times [\chi^{\dagger}(\mathbf{k}_{i},s)
             \chi(\mathbf{k}_{j},s')],
\label{eq:functional}
\end {eqnarray}
%
where $N_{s}$ are the numbers of occupied orbitals with $s=\pm 1$,
$E(\mathbf{k}_{i},s)$ are the HF single-particle energies,
$v(\mathbf{k}_{i}-\mathbf{k}_{j})=2\pi e^{2}/\varepsilon
|\mathbf{k}_{i}-\mathbf{k}_{j}|$ is the Fourier transform of the
Coulomb interaction, and
%
%
\begin {equation}
H_{R}(\mathbf{k}) = \left(
      \begin{array}{cc}
         \hbar^{2}k^{2}/2m^{*} & i\alpha\langle E_{z}\rangle ke^{-i\varphi} \\
                               &                                            \\
                               &                                            \\
         -i\alpha\langle E_{z}\rangle ke^{i\varphi}   &   \hbar^{2}k^{2}/2m^{*}
      \end {array}
\right). \label{eq:matrashba}
\end {equation}
We minimize this functional with respect to the amplitudes
$\chi_{\epsilon}^{*}(\mathbf{k}_{i},s)$ and obtain the
single-particle energies,
%
\begin{eqnarray}
&& E(\mathbf{k},s)=\frac{\hbar^{2}k^{2}}{2m^{*}}+s\alpha \langle
E_{z} \rangle k \sin\theta(\mathbf{k})  \nonumber   \\
&& -\frac{1}{2A}\sum_{s',\mathbf{k'} \in \mathcal{D}_{s'}}
v(\mathbf{k}-\mathbf{k'})\big[1+ss'\hat{\mathbf{u}}(\mathbf{k})\cdot
\hat{\mathbf{u}}(\mathbf{k'})\big].
 \label {eq:singlenergies}
\end {eqnarray}
%
By demanding that the single-particle HF Hamiltonian be diagonal on
the basis of the spinor amplitudes
$\chi_{\epsilon}(\mathbf{k}_{i},\pm)$ we obtain the following
integral  equations:
%
%
%
\begin {eqnarray}
&& 2\alpha \langle E_{z} \rangle k \sin[\phi(\mathbf{k})-\varphi]
\cos\theta(\mathbf{k})=  \nonumber   \\
&&\frac{1}{A}
      \sum_{s',\mathbf{k'} \in  \mathcal{D}_{s'}}
             s'v(\mathbf{k}-\mathbf{k'})
             \{
                   \sin \theta(\mathbf{k})
                   \cos\theta(\mathbf{k'})  \nonumber   \\
&&                   - \sin\theta(\mathbf{k'})
                     \cos\theta(\mathbf{k})
                     \cos[\phi(\mathbf{k})- \phi(\mathbf{k'})]
             \},
\label {eq:nonlinear1}
\end {eqnarray}
%
%
\begin {eqnarray}
&& 2\alpha \langle E_{z} \rangle k \cos[\phi(\mathbf{k})-\varphi] =
\frac{1}{A}
      \sum_{s',\mathbf{k'} \in  \mathcal{D}_{s'}}
             s'v(\mathbf{k}-\mathbf{k'})   \nonumber   \\
&&             \times\sin\theta(\mathbf{k'})\sin[\phi(\mathbf{k})-
             \phi(\mathbf{k'})].
\label {eq:nonlinear2}
\end {eqnarray}
%
In these equations, we have substituted the following expressions
for $\chi_{\epsilon}(\mathbf{k}_{i},\pm)$ in order to display the
dependence on the polar angles $\theta(\mathbf{k})$ and
$\phi(\mathbf{k})$: $\chi_{+}(\mathbf{k},+)=
\cos(\theta/2)\exp(-i\phi/2)$,
$\chi_{-}(\mathbf{k},+)=\sin(\theta/2)\exp(i\phi/2)$,
$\chi_{+}(\mathbf{k},-)=-\sin(\theta/2)\exp(-i\phi/2)$, and
$\chi_{-}(\mathbf{k},-)=\cos(\theta/2)\exp(i\phi/2)$. The summation
domains $\mathcal{D}_{\pm}$ are the regions of $k$ space occupied by
electrons and their areas are $N_{\pm}$, respectively. Notice that
Eq.\ (\ref{eq:singlenergies}) reduces to: (i) Eq.\
(\ref{eq:energirash}) when the Coulomb interaction is neglected
because the third term of the right-hand side (RHS) of Eq.\
(\ref{eq:singlenergies}) drops out and $\theta(\mathbf{k})=\pi /2$
in the single-particle Rashba problem; (ii) the HF one-particle
energy if the Rashba coupling is omitted because $\alpha =0$ and
$\hat{\mathbf{u}}(\mathbf{k})\equiv\hat{\mathbf{z}}$.

\section{\label{sec:circular}isotropic solution}

In the absence of Pomeranchuk instabilities \cite{pom} (PI)
(deformations of the Fermi sphere in three dimensions or circle in
two dimensions), the domains $\mathcal{D}_{\pm}$ should be taken as
having circular symmetry.
The issue of the occurrence of PI in Fermi liquids with isotropic
central interactions is currently being studied in two-dimensional
(2D) and three-dimensional (3D) systems.\cite{qui-sch,qui-hoo-etal}
In the particular case of the bare Coulomb interaction (which is our
case), the existing theory is not able to categorically predict or
rule out the PI. However, a screened Coulomb interaction does not
produce PI (Ref.\ \onlinecite{qui-sch}), which can be taken as an
indication that no PI occurs either for the bare interaction. In
general, it is safe to assume that no PI will occur unless: (i)
there is a well-defined length scale in the interaction, and (ii)
that length scale is larger than the mean interparticle
distance.\cite{qui-pri} These conditions are clearly not satisfied
by the bare Coulomb interaction, which does not possess a
characteristic length scale. Based on these results, it is safe to
assume that our integration domains have circular symmetry.

The circular symmetry of the integration domains implies that the
dispersion relations $E(\mathbf{k},s)$ in Eq.\
(\ref{eq:singlenergies}) must be isotropic. This in turn, requires
that $\theta(\mathbf{k})$ be independent of $\varphi$ and that
$\phi(\mathbf{k}) = \varphi - \pi/2$ as in the noninteracting Rashba
problem.\cite{footfi} Thus Eq.\ (\ref{eq:nonlinear2}) is
automatically satisfied and Eq.\ (\ref{eq:nonlinear1}) becomes
%
%
%
\begin {eqnarray}
px\cos\theta(x)=\int_{x_{c}}^{1}\int_{0}^{2\pi}\frac{x'dx'd\varphi
'}{\sqrt{x^{2}+x'^{2}-2xx'\cos(\varphi-\varphi')}}    \nonumber   \\
\times[\sin\theta(x)\cos\theta(x')-\cos\theta(x)\sin\theta(x')\cos(\varphi-\varphi')].
 \label {eq:ecinteg}
\end {eqnarray}
%
We have introduced the parameter $p = 2 \alpha \langle E_z \rangle
\varepsilon/e^2$ which indicates the relative strength of the Rashba
and Coulomb interactions. The integration limit $x_c$ contains the
information of the integration domains $\mathcal{D}_{\pm}$
introduced earlier. Having in mind the dispersion relation of the
noninteracting Rashba problem described in Eq.\
(\ref{eq:energirash}), the situation can be summarized as follows:
when both branches are occupied (high density), the domains
$\mathcal{D}_{\pm}$ are filled Fermi circles of radii $k_{F\pm}$
related by $4\pi n_{s}=k_{F+}^{2}+k_{F-}^{2}$, with
$n_{s}=(N_{+}+N_{-})/A$.
In this case $x_c = k_{F+}/ k_{F-}$ and, in principle, it is free to vary
from zero to one.
If only the lower branch is occupied (low density), there are two
possibilities.
If the lower branch has a minimum at $k=0$ (unlike the
noninteracting Rashba problem), then there is a gap at $k=0$ between
the two branches.
In this case we define $x_c$ as before, obtaining $x_c=0$.
The formation of a gap is enabled by the lifting of Kramers degeneracy
at the single-particle level, due to the appearance of a spontaneous
magnetization.
Kramers' theorem states that the degeneracy of half-integer spins
can only be removed by a magnetic field; we return to this point in
the discussion of Eq.\ (\ref{eq:gap}).
If the lower branch does have a minimum at $k\neq 0$ we take
$x_c = k_{min} / k_{max}$, where
$k_{min}$ $(k_{max})$ are the inner (outer) Fermi radii of the
hollow circular domain, now being $4\pi
n_{s}=k_{max}^{2}-k_{min}^{2}$.
In this case, there is no gap at $k=0$, no (ground-state) magnetic moment
appears and the time-reversal symmetry (Kramers degeneracy) is preserved.
In Eq.\ (\ref{eq:ecinteg}), $x= |\mathbf{x}| \equiv |\mathbf{k}| /
k_{F-}$ or $x= |\mathbf{k}| / k_{max}$, according to the context.

Clearly, the noninteracting Rashba states with $\theta = \pi /2$,
are a solution of Eq.\ (\ref{eq:ecinteg}). We call this the in-plane
(IP) paramagnetic phase, which is the one we mentioned as having the
lower dispersion-relation branch with a minimum at $k\neq 0$. The
nontrivial solution with varying $\theta(x)$ gives rise to an OP
ferromagnetic phase where, in turn, the lower branch has its minimum
at $k=0$. We solve Eq.\ (\ref{eq:ecinteg}) starting with the initial
guess of $\theta _{0}(x)=0$, which gives $\theta _{1}(x)$ after
integration. With $\theta _{1}(x)$ as an input, we obtain $\theta
_{2}(x)$ and so on. We consider that convergence is achieved when
$\theta _{n}(x)$ and $\theta _{n-1}(x)$ differ in less than $0.1$
percent. Let us enumerate our main findings on $\theta(x)$: (i)
$\theta(0)=0$ for all values of $p$ (the Rashba spin-orbit
interaction causes no effect when $k=0$ and consequently, the spin
quantization axis must lie in the $z$ direction), (ii) $\theta(x)$
is a monotonically increasing function for all values of $p$, (iii)
$\theta(x)$ increases with increasing $p$, and (iv) $\theta(x)$
never crosses the value $\pi/2$.

The gap at $k=0$ between the two branches may be obtained from Eq.\
(\ref{eq:singlenergies}) by means of the above-mentioned properties,
and it is given by
\begin {equation}
\Delta
E=E(0,+)-E(0,-)=\frac{e^{2}k_{F-}}{\varepsilon}\int_{xc}^{1}\cos
\theta(x)dx. \label {eq:gap}
\end {equation}
In the IP phase $\theta(x)=\pi/2$ and then $\Delta E=0$. In the OP
phase, in turn, we have $\Delta E\neq 0$. This gap, far from being a
peculiarity of the OP phase, appears also in the HF theory of the
2DEG without Rashba coupling. In fact, setting $\theta (x) = 0$ we
get $\Delta E=(e^{2}/\varepsilon)(k_{F-}-k_{F+})$ and the gap is
unfailingly related to a polarized ground state, where $k_{F+}\neq
k_{F-}$ (i.e.,\ $N_{+}\neq N_{-}$) (an identical expression for
$\Delta E$ multiplied by $2/\pi$ can be obtained in a 3DEG: see the
expression for $\epsilon^{\pm}_{k}$ when $k\rightarrow 0$ on p.\ 82
of Ref.\ \onlinecite{gro-run-hei}).

\section{\label{sec:ground}Hartree-Fock ground state}

\begin{figure}
 \scalebox{.23}
    {\includegraphics {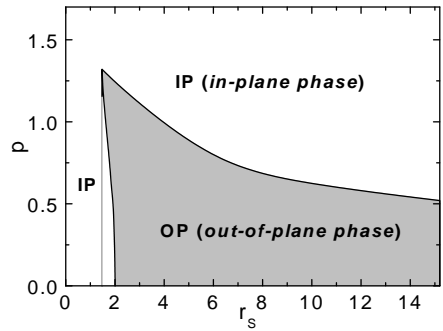}}
    \caption{\label{fig1}
    Ground-state phase diagram in the Hartree-Fock approximation in
    terms of the density parameter $r_s$ and the Rashba to Coulomb
    energy ratio $p$.}
\end{figure}

The remaining task is, for given $p$ and $r_s$
($r_{s}=1/a_{B}^{*}\sqrt{\pi n_{s}}$, where
$a_{B}^{*}=\hbar^{2}\varepsilon/m^{*}e^{2}$ is the effective Bohr
radius), to determine which of these two phases has lower energy.
For each phase and given $r_{s}$, the value of $x_{c}$ that
minimizes the energy is found numerically. In \mbox{Fig.\
\ref{fig1}} we show a phase diagram in terms of the parameters $r_s$
and $p$. The striking feature of this diagram is that for given
$0<p\lesssim 1.3$, the OP phase appears within a window of
densities. As expected, for $p=0$ we recover the
paramagnetic-ferromagnetic transition of the 2DEG HF approximation
at $r_s=2.01$ (Refs.\ \onlinecite{rad-tam-das} and
\onlinecite{kim-raj}). As $p$ increases, the \textit{left}
transition moves slightly towards smaller $r_{s}$; in other words,
the presence of the Rashba coupling favors a spin-polarized phase,
albeit this polarization is partial for nonzero $p$ (the original HF
ferromagnetic phase has full polarization). Also the window of
densities in the OP phase shrinks as $p$ increases. The
\textit{right} transition originates in the fact that the system
diminishes the ground-state energy by filling the lower branch that
has minimum at $k\neq 0$ (IP phase), when $r_{s}$ increases at fixed
$p$.

%
%

\begin{figure}
 \scalebox{.23}
    {\includegraphics {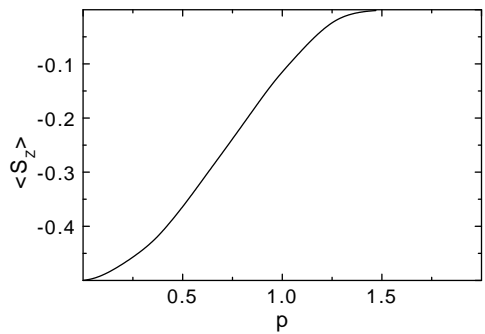}}
    \caption{\label{fig2} Mean value of the perpendicular spin
     projection per particle ($\hbar =1$) as a function of the Rashba to Coulomb
     energy ratio in the OP phase.}
\end{figure}

The partial spin polarization of the (ferromagnetic) OP phase can be
seen in \mbox{Fig.\ \ref{fig2}}, where we plot the mean value of
$S_z$ per particle ($\langle S_{x}\rangle =\langle S_{y}\rangle=0$
for both IP and OP phases) given by ($\hbar = 1$),
%
%
%
%
%
\begin {equation}
    \langle S_{z}\rangle =-\int_{x_{c}=0}^{1}  x\cos\theta(x)dx .
    \label{eq:spinprom}
\end {equation}
Notice that $\langle S_{z}\rangle$ does not depend on $r_{s}$ since
$x_{c}=0$ for the OP phase.\cite{footxc}

\section{\label{sec:exper}comparison with experiment and conclusion}

In a recent experimental work, Ghosh \textit{et al.}~\cite{gho}
report a possible spontaneous spin polarization in mesoscopic
two-dimensional systems.
They studied 2DEGs in asymmetric Si $\delta$-doped GaAs/AlGaAs heterostructures
with densities as low as $n_s = 5\times10^{9}$cm$^{-2}$ ($r_{s}=7.6$).
The temperature was set at $T \approx 40$ mK or equivalently $T/T_{F}\approx
0.02$ since $T_{F}=2.3$ K at $r_{s}=7.6$.
According to their interpretation of the data, these authors found
partial spin polarization estimated as $\zeta\equiv (N_{+}-N_{-})/N
\approx 0.2$ appearing in a window of densities of width $\Delta
r_{s}\approx 1.8$, centered around $r_{s}\approx 6.5$.
After a detailed analysis of their data, the authors rule out the Rashba
coupling and invoke an exchange-driven spontaneous spin polarization
in order to explain their observed split zero-bias peak (ZBP) in the
differential conductance.
On the other hand, they explain the fact that the spin polarization
is partial as a finite-temperature effect.
Recall that at zero temperature, this transition is of first order
and is predicted to occur between $r_{s}\approx 13$ (Ref.\
\onlinecite{cep}) and $r_{s}\approx 25$ (Ref.\ \onlinecite{att}).
However, this contradicts the theoretical finding of Dharma-wardana
and Perrot, \cite{dha1} in the sense that partial spin polarization
due to the exchange happens for $T/T_{F}$ between 0.3 and 1.6,
i.e.,\ well above $T/T_{F}\approx 0.02$ as reported by Ghosh
\textit{et al}.

Our theory supports the interpretation of the experiment of
Ghosh \textit{et al.} based on an exchange-driven spin-polarized ground state.
On the other hand, we explain the partial spin polarization in terms
of the Rashba interaction rather than in terms of finite temperature.
This hypothesis could be tested by additional experiments with a
tunable Rashba coupling, which would allow varying our parameter $p$.
In such experiments the degree of spin polarization could be modified
and measured.

Ghosh {\it et al.}\ observe a split ZBP within a window of electron
densities.
At the lower end of this window, disorder-induced localization destroys
the electron-gas scenario.
In a defect-free sample, we also predict a disappearance of the
split ZBP but due entirely to the competition between Rashba and
exchange interactions.\cite{footwindow}
At the higher end of the density window, the reason for the disappearance
of the split ZBP is not as clear, and it is not specified by the authors.
According to our theory, this feature could be interpreted as the
ferromagnetic-paramagnetic transition of the electron gas at high
density, seen in Fig.\ \ref{fig1} for low $r_s$. (Of course, the
critical value of $r_s$ is overestimated by the Hartree-Fock
approximation because it leaves out many-body correlations.
\cite{jur-tam})

In summary, we have found the uniform Hartree-Fock ground state of a
2DEG in the presence of spin-orbit Rashba coupling, characteristic
of asymmetric semiconductor quantum wells.
We present a phase diagram where two competing quantum phases are
identified, one of which shares the Rashba single-particle orbitals,
with their IP spin quantization axis, and another one which has a
finite component of the spin in the perpendicular direction.
This phase possesses a partial spin polarization and exists within a
window of densities, suggesting that a combination of exchange and
Rashba spin-orbit interaction may qualitatively explain experimental
results obtained in GaAs-AlGaAs heterostructures at low electron
density.

\begin{acknowledgments}
We thank J.\ Quintanilla for his useful comments. We acknowledge
financial support from UBACyT, CONICET and ANPCyT. P.I.T.\ is a
researcher of CONICET.
\end{acknowledgments}


\begin{thebibliography}{99}
\bibitem{footfi}
More precisely, the circular symmetry of the dispersion relation
requires that $\phi(\mathbf{k})=\phi_{0}(k) + n\varphi$ where $n$ is
an integer and $\phi_{0}(k)$ is an arbitrary function. However, Eq.\
(\ref{eq:nonlinear2}) determines that $n=1$ and analyzing the
expression for the HF ground-state energy (not shown), we realize
that it is a minimum when $\phi_{0}(k)=\phi_{0}=const$. Finally,
Eq.\ (\ref{eq:nonlinear2}) forces $\phi_{0}=\pm \pi /2$, and as in
the single-particle Rashba problem, we choose $N_{-}$ as the
majority number, i.e.,\ $\phi_{0}=-\pi /2$.
\bibitem{footxc}
The $r_{s}$ dependence of $x_{c}$ may be summarized as follows: (i)
in the OP phase $x_{c}=0$ because only the lower branch is occupied,
(ii) in the IP phase at high density, on the contrary, for given
$p$, $x_{c}$ decreases monotonically as $r_{s}$ increases due to the
occupancy of both branches, and (iii) $x_{c}$ increases
monotonically in the IP phase at low density since only the lower
branch is occupied and $k_{min} \rightarrow k_{max}$ when $r_{s}$
increases.
\bibitem{footwindow}
Thus, localization at low density anticipates and masks the
OP-IP transition seen in Fig.~\ref{fig1}.
%
Note from Fig.~\ref{fig2} that $\zeta \approx 0.2$
corresponds to $p\approx 1$ and with this value we get, from
Fig.~\ref{fig1}, a window of densities $\Delta r_{s}\approx 2$.
\end{thebibliography}
\end {document}